\documentclass{article}
\usepackage{graphicx}

\begin{document}

\centerline{\Large \bf Lenticular galaxies with UV-rings}
\medskip
\centerline{\large \it M.A. Ilyina, O.K. Sil'chenko}
\medskip
\centerline{\it Sternberg Astronomical Institute of the MSU, Moscow}

\bigskip
\bigskip

{\bf
By using the public UV imaging data obtained by the GALEX (Galaxy Ultraviolet
Explorer) for nearby galaxies, we have compiled a list of lenticular galaxies 
possessing ultraviolet rings -- starforming regions tightly confined to 
particular radial distances from galactic centers. We have studied large-scale
structure of these galaxies in the optical bands by using the data of the
SDSS (Sloan Digital Sky Survey): we have decomposed the galactic images into
large-scale disks and bulges, have measured the ring optical colours from
the residual images after subtracting model disks and bulges, and have
compared the sizes of the rings in the optical light and in the UV-band.
The probable origin of the outer starforming ring appearances in unbarred 
galaxies demonstrating otherwise the regular structure and homogeneously old 
stellar population beyond the rings is discussed.
}
\bigskip
\bigskip

\section{Introduction}

According to the Hubble's classification, lenticular galaxies, as an intermediate
class between ellipticals and spiral galaxies, possess large-scale stellar
disks but lack noticeable starforming sites. This criterium refers to the
optical-band images of galaxies. However the recent morphological survey of
nearby galaxies in the UV-bands undertaken by the GALEX space telescope
\cite{galex} has demonstrated that star formation is present
in lenticular galaxies more often than it has been thought before, and
a typical pattern of the star formation in lenticular galaxies is a large-scale 
ring.

Large-scale rings in disk galaxies are well-known structures; their classification
is established long ago and is related to their origin. Few and Madore \cite{few}
analysed a sample consisting of 69 ring galaxies and divided all the rings into two
types -- `O-rings' and `P-rings'. The former class included regular smooth rings
with the galactic nuclei in their centers, the latter type had inhomogeneous
distributions of surface brightness along the rings and the geometrical centers
displaced often with respect to the galactic nuclei. The counts of `satellites' --
smaller galaxies around the hosts of the rings -- proved that `P-rings' are
probably collisional structures formed after the satellite crossing the main
galaxy disks. The `O-rings' did not relate necessarily to galaxy interaction;
they might be resonance structures. Schwarz \cite{schwarz} considered the response of
a large-scale gaseous disk to a bar (non-axisymmetrical distortion of the disk
gravitational potential) and found that in course of its dynamical evolution 
the gas had to accumulate in a ring located at the outer Lindblad resonance of 
the bar. Evidently, the gas accumulation there could lead to a star formation burst
in the ring at this particular radius. Buta and Crocker \cite{buta} analysed 
observational statistics of metric sizes of the rings for a sample of a few hundred 
galaxies and confirmed that the rings were related to bar resonances. However,
several unbarred galaxies are also known that have regular starforming rings;
every such case requires individual analysis after which a conclusion follows
usually that the galaxy has experienced a minor merger having produced this
ring \cite{we_rings}.

After the results of the nearby galaxy survey in the UV-bands by the GALEX 
have become available, the researchers have reported outer UV-rings in several
regular lenticular galaxies lacking star formation signatures in the optical
bands. The UV-rings have been noted in IC 4200 \cite{ic4200}, NGC 2962 
\cite{marino}, ESO 381-47 \cite{donovan}, NGC 4262 \cite{n4262uv},
and in NGC 404 \cite{thilker10}; they may have either collisional
or resonance origin, with various probability degree in every particular
case. We think that the very rise of a starforming ring in a lenticular 
galaxy otherwise lacking significant gas supply or young stars may be related
to its transformation from a spiral galaxy. Here we present the first 
sample of regular lenticular galaxies with the UV-rings which can be
used for a statistical analysis, and discuss possible origin of these rings
in minor mergers. To exclude evidently resonance-triggered rings which
do not need galaxy interaction for their formation, we take only unbarred
galaxies.

\section{The sample}

Firstly, we have looked at the UV-images of the nearby galaxies included
into the Atlas by Gil de Paz et al. \cite{galex} and have identified the galaxies
with the ring morphology in the UV-bands. Only the galaxies observed also
by the SDSS are included into our sample because we want to have a possibility 
to compare the UV and optical morphologies of our targets. Secondly, we have 
used the list of ringed galaxies compiled by Kostyuk \cite{ira} by inspecting the
Palomar Atlas images, such that these galaxies have outer rings in the optical
blue band. For the unbarred S0-galaxies from the Kostyuk's list we have retrieved
the GALEX images by exploring facilities of the NED links and have found several
UV-bright outer rings among these galaxies too.

For all S0-galaxies found to possess UV-rings we have retrieved $g$ 
($\lambda$4686~\AA) and $r$ ($\lambda$6165~\AA) digital images from the
public data archives of the SDSS, DR7 \cite{sdssdr7} and DR8
\cite{sdssdr8}, to search for the rings in the optical bands and
to compare the properties of the `optical' rings with those in the UV.

The final sample includes 14 S0-galaxies; their global properties taken from
the extragalactic databases are listed in the Table~1.

{\scriptsize
\begin{table*}
\caption[ ] {Global parameters of the galaxies under consideration}
\begin{flushleft}
\begin{tabular}{lccccccccl}
\hline\noalign{\smallskip}
Galaxy  & Type  & $V_r$ (km/s)  & D(Mpc),$^2$ &
$R_{25},\, ^{\prime \prime}$   & $m_B$  & $M_B$ 
& $u-r$ & HI? & Environment  \\
 & NED$^1$ & NED & NED & RC3$^3$ & RC3 & LEDA$^4$ & NED & NED & NED \\
\hline
IC~522 & S0 & 5079 & 71 & 30 & 13.97 & --20.8 & 2.75 & -- & single$+$satellite \\
MCG~11-22-15 & -- & 8064 & 114 & 24$^4$ & 15.6$^4$ & --19.9 & -- & -- &
a group member \\
NGC~252 & (R)SA0$^+$(r) & 4938 & 71 & 45 & 13.35 & --21.3 & 0.97$^6$ & $+$ & a group
center \\
NGC~446 & (R)SAB0$^0$ & 5446 & 76.5 & 61 & 13.35 & --21.0 & -- & $+$ & in a pair \\
NGC~809 & (R)S0$^+$: & 5367 & 74 & 44 & 14.66$^5$ & --19.9 & 2.69 & -- &
single$+$3 sat \\
NGC~934 & SAB0$^-$ & 6353 & 88 & 40 & 14.04 & --20.7 & 2.69 & -- & a group member \\
NGC~4344 & SB0: & 1142 & 14.5 & 50 & 13.34 & --18.2 & 2.02 & $+$ & a Virgo member \\
NGC~4513 & (R)SA0$^0$ & 2304 & 34 & 43 & 14.01 & --19.0 & 2.61 & $+$ & in a wide pair \\
NGC~6028 & (R)SA0$^+$: & 4475 & 62.5 & 40 & 14.35 & --20.0 & 2.76 & $+$ &
a group center \\
NGC~6340 & SA0/a(s) & 1198 & 20 & 97 & 11.87 & --20.0 & 2.78 & $+$ & a group
center \\
NGC~6534 & S? & 8332 & 117.5 & 25 & 15.40 & --20.1 & -- & -- & isolated \\
NGC~7808 & (R')SA0$^0$: &  8787 & 122 & 38 & 13.48 & --21.3 & 2.59 & -- &
in a pair$+$5 sat \\
UGC~4599 & (R)SA0$^0$ & 2072 & 26 & 60 & 13.6 & --17.5 & 2.45 & $+$ & a group
center \\
UGC~5936 & (R)SA0$^+$: & 7230 & 99 & 39 & 14.21 & --21.0 & 2.66 & -- &
in a triplet \\
\hline
\multicolumn{10}{l}{$^1$\rule{0pt}{11pt}\footnotesize
NASA/IPAC Extragalactic Database}\\
\multicolumn{10}{l}{$^2$\rule{0pt}{11pt}\footnotesize
distance from the NED, with respect to the Local Group}\\
\multicolumn{10}{l}{$^3$\rule{0pt}{11pt}\footnotesize
Third Reference Catalogue of Bright Galaxies \cite{rc3}}\\
\multicolumn{10}{l}{$^4$\rule{0pt}{11pt}\footnotesize
Lyon-Meudon Extragalactic Database}\\
\multicolumn{10}{l}{$^5$\rule{0pt}{11pt}\footnotesize
the blue magnitude $b_j$, from the APM \cite{maddox}}\\
\multicolumn{10}{l}{$^6$\rule{0pt}{11pt}\footnotesize
it is a colour $B-V$ instead of $u-r$ which is absent in the NED}\\
\end{tabular}
\end{flushleft}
\end{table*}
}

\section{Surface photometry in the optical bands}

Our processing with the optical images included:

\begin{itemize}
\item{constructing $g-r$ colour maps for the initial (full) galaxy images;}
\item{decomposing every galaxy into a large-scale Sersic bulge and exponential
disk(s) in $g$-band and in $r$-band independently;}
\item{deriving residual images by subtracting the obtained model images 
bulge$+$disk(s) from the initial (full) galaxy images;}
\item{constructing $g-r$ colour maps for the residual images.}
\end{itemize}

All the calculations besides the decomposition have been made with the software
by V.~Vlasyuk \cite{vlasyuk}; the photometric calibration information has been
taken from the SDSS WEB-site. The decompositions, and also the colour profiles
for the residual images, were made with the software GIDRA \cite{moisav}.
To fit the initial (full) images of the galaxies, we applied a model consisting
of two exponential disks with different scalelengths and of a Sersic bulge. 
With such models, all the galaxies except NGC~4344 were modelled quite finely.
To describe the surface brightness profiles of NGC~4344, only two exponential
segments are enough; taking into account also the inner structure of the galaxy,
we conclude that probably NGC~4344 has no bulge at all. The results of the
decompositions in the $g$- and $r$-bands for all galaxies of the sample which are
characterized by the central brightnesses and Sersic-law scalelengths are presented 
in the Table~2.

{\scriptsize
\begin{table*}[h!]
\caption{Photometric parameters of the galaxy decompositions}
\begin{flushleft}
\begin{tabular}{|r|c|cc|cc|ccl|}
\hline
Galaxy & Band & 
\multicolumn{2}{|c|}{Outer disk} & 
\multicolumn{2}{|c|}{Inner disk} & \multicolumn{3}{|c|}{Bulge} \\
&  & $r_0$, $^{\prime \prime}$   &
$\mu_0$, mag/$arcsec^2$ &
$r_0$, $^{\prime \prime}$  &
$\mu_0$, mag/$arcsec^2$&
 $r_0$, $^{\prime \prime}$ &
$\mu_0$, mag/$arcsec^2$ & $n$ \\
\hline
	 	     
	IC~522 & $r$ & 27.8 & 23.4 & 5.9 & 19.3 & 2.37 & 12.7 & 1.0 \\ 
	IC~522 & $g$ & 22.3 & 23.6 & 5.8 & 20.1 & 2.26  & 13.5 & 1.0 \\
      MCG~11-22-15 & $r$ & 10.7 & 26 & 5.2 & 23.2 & 1.51 & 15.5 & 1.9   \\	
	MCG~11-22-15 & $g$ & 11.0 & 25.5 & 4.9 & 24.3 & 1.54 & 16.4 & 1.9   \\
      NGC~252 & $r$ & 73 & 24.5 & 14.9 & 20.5 & 3.73 & 17.4 & 1.5   \\
	NGC~252 & $g$ & 58 & 24.7 & 12.1 & 21.2 & 4.13 & 18.9 & 1.6   \\	
      NGC~446 & $r$ & 23.9 & 24.9 & 11.5 & 22.7 & 3.32 & 14.5 & 1.8  \\				
	NGC~446 & $g$ & 23.3 & 25.4 & 12.1 & 23.3 & 2.9 & 15.5 & 1.8   \\
	NGC~809  & $r$ & 18.0 & 23.8 & 10.2 & 19.5 & 1.4 & 17.2 & 1.2 \\ 
	NGC~809  & $g$ & 20.3 & 23.1 & 10.3 & 20.5 & 2.33 & 15.8 & 1.4 \\
	NGC~934  & $r$ & 35 & 23.0 & 7.3 & 20.7 & 2.24 & 15.3 & 2.1 \\ 
	NGC~934  & $g$ & 42.6 & 24.3 & 7.0 & 21.4 & 2.01 & 16.6 & 1.8 \\ 
	NGC~4344 & $r$ & 47.1 & 22.9 & 10.3 & 19.8 & -- & -- & -- \\
	NGC~4344 & $g$ & 62.5 & 24.4 & 10.7 & 20.2 & -- & -- & -- \\
      NGC~4513 & $r$ & 79.3 & 23.7 & 8.9 & 21.3 & 1.97 & 15.8 & 1.5   \\	
	NGC~4513 & $g$ & 63 & 24 & 10.1 & 22.3 & 2.15 & 16.5 & 1.6   \\	
	NGC~6028 & $r$ & 25 & 21.8 & 10.0 & 19.5 & 1.77 & 14.3 & 2.2   \\
	NGC~6028 & $g$ & 25 & 22.4 & 12.4 & 21.8 & 1.66 & 15.1 & 2.2   \\
	NGC~6340 & $r$  & 50.4 & 22.2 & 20.3 &  20.0 & 5.17 & 15.1 & 1.9   \\ 
	NGC~6340 & $g$  & 73.3 & 22.9 & 22.1 &  20.8 & 5.23 & 15.3 & 2.0   \\ 
      NGC~6534 & $r$ & 15.7 & 21.3 & 7.6 & 20.2 & 1.95 & 17.2 & 1.8   \\		
	NGC~6534 & $g$ & 14.2 & 22.9 & 6.9 & 20.7 & 2.31 & 18.2 & 1.8   \\	
	NGC~7808 & $r$ & 46 & 24.3 & 7.6 & 20.2 & 2.66 & 16.2 & 1.7   \\
	NGC~7808 & $g$ & 63 & 24.9 & 8.6 & 21.3 & 2.93 & 17.3 & 1.7   \\ 
	UGC~4599 & $r$ & 56.4 & 23.8 & 5.5 & 19.9 & 2.24 & 16.5 & 2.1   \\	
	UGC~4599 & $g$ & 50.1 & 24.3 & 5.1 & 20.3 & 1.59 & 17.2 & 2.1   \\		
	UGC~5936 & $r$ & 21.4 & 24 & 7.7 & 21.9 & 1.82 & 16.1 & 1.5   \\	
	UGC~5936 & $g$ & 23.4 & 24.8 & 7.3 & 22.6 & 1.82 & 16.9 & 1.5   \\	
\hline
\end{tabular}
\end{flushleft}
\end{table*}
}

\section{The results: ring parameters}

The general conclusions which we can made from our calculations are the following:

\begin{itemize}
\item{if to exclude NGC~809 and MCG~11-22-15 which have red rings, the rings of 
all other galaxies look rather blue in the visible light;}
\item{the radii of the rings, their location in the galaxies are the same in the
UV rays and in the optical bands.}
\end{itemize}

The main parameters of the rings estimated by using the residual images in the
optical bands are given in the Table~3. The colours of the rings are measured
at the middle radii of the rings; only for UGC~5936 and MCG 11-22-15 we give
also the colours at the inner and outer edges of the rings because they differ
substantially from the colours at the middle radii: in the middle, the rings
in these two galaxies are very blue, and they become redder close to the edges. 
Let us to discuss the results for the individual galaxies.\\

{\scriptsize
\begin{table*}[h!]
\caption{The characteristics of the rings in the UV and in the optical bands} 
\begin{flushleft}
	\begin{tabular}{||r|c|c|c||}
	\hline
	Galaxy &   Optical radii (arcsec) &  UV-radii (arcsec)  & $g$-$r$ \\ 
\hline	 	     
	IC~522	&	11-24		& 13-26		&		0.56 \\ 
      MCG~11-22-15 &	10-22		&	14-21	&	0.90(middle) 1.00(edges)\\
      NGC~252 &	21-30		&	22-34		&	0.30		\\
      NGC~446 &	25-42		&	25-42		&	0.84		\\
	NGC~809  &   16-34     &  7-32      &       1.15 \\ 
	NGC~934  &  about 57-80   &  50-88  &      0.37 \\ 
	NGC~4344 &    3-12       &  2.2-11.5  &     0.35 \\ 
      NGC~4513 &	about 60-80		&	69-83		&	0.70		\\
      NGC~6028 &	23-40		&	19-39		&	0.60		\\
	NGC~6340 &    47-64;25-35;4-20     &  up to 66  & 0.69; 0.41; 0.46  \\ 
	NGC~6534 &	10-21		&	11-22		&	0.30		\\	
	NGC~7808 &   18-34      &    23-36    &      0.65                 \\ 
	UGC~4599 &	39-60		&	41-58	&	0.40	\\	
	UGC~5936 &	14-38		&	18-33	&	0.64(middle) 0.83(edges)\\
	\hline
	\end{tabular}
\end{flushleft}	
	\end{table*}
}

\textit{IC~522}. The galaxy is decomposed into two disks and a Sersic bulge
with the exponential brightness profile. After subtracting this model from the
galaxy images, we see one more centrally concentrated component in the
residuals; due to its red, round appearance, this component may be a small 
de Vaucouleurs' bulge inside the more extended pseudobulge. The ring looks
asymmetric both in the UV and in the optical bands. The galaxy can be treated
as practically isolated: acording to the NED, it has only one faint satellite.\\

\textit{MCG 11-22-15}. The decomposition of the optical images has enlightened
the presence of the regular ring and a minibar in the center. The galaxy resembles
UGC~5936 by its residual structure; the only difference is the more narrow 
optical ring extension in the north-western segment of the residual images.
We cannot estimate if the UV-ring has a similar squeeze because the spatial
resolution of the GALEX images is insufficient. Also we would note the rather
red optical colour of the ring in this galaxy.\\

\textit{NGC~252}. The outstanding peculiarity of the ring is this galaxy is
that it is obviously inclined to the main symmetry plane of the galaxy. In
the central part of the residual images a spiral pattern is seen that is
an argument in favour of the circumnuclear disk presence. The colour
residual map demonstrates a very blue $g-r$ of the ring.\\

\textit{NGC~446}. The ring is among the reddest ones with its optical colour
of $g-r=0.84$. The galaxy is rich with diffuse matter that is seen both in
the optical and UV images. Also the residual images reveal some structure
in the center of the galaxy which may be identified as a circumnuclear disk
with a spiral pattern.\\

\textit{NGC~809}. The residual images of this galaxy contain a lot of interesting
details. Besides the very red, probably dust-rich ring, we must note the
central minibar with weakly developped spiral arms. The UV- and optical rings
demonstrate comparable brightness contrasts.\\ 

\textit{NGC~934}. While the UV-ring of the galaxy is bright, though 
slightly asymmetric, the optical ring is less impressive and can hardly 
be distinguished against the noise background of the residual images.
Instead, a bright circumnuclear disk is seen both in the UV and in the optical
light. The galaxy is a member of a rather rich group.\\

\textit{NGC~4344}. As the surface brightness profile of this galaxy
implies, NGC~4344 lacks a bulge that is quite unusual for a lenticular
galaxy. The ring is strongly inhomogeneous, with bright star forming knots
visible both in the UV and in the optical bands. The residual images look
asymmetric, with the south-eastern part more rich in matter than the north-western
one. The galaxy belongs to the Virgo cluster.\\

\textit{NGC~4513}. The outer ring of this galaxy resembles that of NGC~934: it is
bright in the UV and faint in the optical bands. The residual $g-r$ colour 
map reveals another blue ring in the inner part of the galaxy -- it cannot 
be distinguished at the GALEX maps because it is deeply embedded into the UV-bright circumnuclear disk.\\

\textit{NGC~6028}. The ring of the this galaxy demonstrates a tail-like extension
at the south-eastern edge. By the decomposition we have put into evidence a presence
of a nuclear minibar. After considering the colour map of the residual image we have
concluded that the object seen to the south from the galactic nucleus is probably 
a foreground star.\\

\textit{NGC~6340}. The galaxy is already studied well \cite{me6340,chil6340}.
The result of our analysis is that it possesses 
at least three asymmetric rings which are brighter in the optical bands than
in the UV. The galaxy locates at the center of a rather rich group.\\

\textit{NGC~6534}. The galaxy has two rings. In the optical light the outer ring
is blue, it is also seen in the UV; the inner ring is rather red and embedded into 
the bulge area. The UV-analog of the inner ring cannot be distinguished because of the lower spatial resolution of the GALEX data. Probably, the two rings are inclined
to the symmetry plane of the galaxy under different angles.\\

\textit{NGC~7808}. By considering the residual maps of this galaxy, 
we have discovered a nuclear minibar and another extended structure co-axial 
with the bar -- perhaps, a circumnuclear stellar disk. The ring which
is also seen at the residual colour map looks asymmetric and rather faint
both in the UV and in the optical light. The galaxy belongs to a small
group consisting of only two large galaxies surrounded by small satellites.\\

\textit{UGC~4599}. The rings of this galaxy, the inner one being similar to that
of NGC~4513 -- clearly visible in the optical colour map and looking only a part of
the bright circumnuclear disk in the UV, -- are rather irregular and consist
of several separate arcs. After subtracting the model of `a Sersic bulge plus two
exponential disks' from the initial galaxy images in the $g$ and $r$ bands,
we have found a small ($2^{\prime \prime}$), red, round structure in the
residual maps in the very center of the galaxy.

\textit{UGC~5936}. The galaxy has a very regular ring, one of the most impressive
in our sample. It resembles somewhat the ring in NGC~809. The residual maps
reveal a presence of the nuclear minibar. The star-like object in the 
south-eastern part of the ring does not belong to the galaxy with a high
probability.

\section{Discussion: the scenaria of ring formation and evolution}

The galaxies of our sample are of S0-type, so first of all, they all possess 
large-scale stellar disks and Sersic bulges, with a power parameter $n\approx 2$
(see the Table~2, the last right column) that is typical for early-type
disk galaxies \cite{graham}. And secondly, they almost all have in general rather old 
stellar populations -- because the integrated colour of $u-r$ is larger than 
2.22 in all galaxies except NGC~4344 (see the Table~1), and this criterium, 
according to Strateva et al. \cite{sdss_col}, attributes them to the `red sequence'.
The first property enables us to separate our galaxies with the outer 
starforming rings from the Hoag object which has a de Vaucouleurs' spheroid -- 
probably an elliptical galaxy in the center of its ring. We would like
to stress this difference because some authors -- Wakamatsu \cite{n6028} studying 
NGC~6028 and Finkelman and Brosch \cite{u4599} studying UGC~4599 -- called them
`Hoag-like galaxies'; we think that it was wrong.

The galaxies of our sample are homogeneously distributed over all types
of environment: they can belong to the field or to groups, among them there
are two quite isolated objects and one Virgo member. The half of the sample
have been reported to possess neutral hydrogen (see the Table~1); this fraction
is in agreement with the known frequency of detecting lenticulars in
the 21~cm line \cite{eder}. However, the more attentive look at the
Table~1 reveals that the galaxies detected in the HI line inhabit the part of
the sample nearest to us, so the visible absence of gas in the other half
of the sample may be a selection effect.

Our sample includes only unbarred galaxies (see the morphological types in
the Table~1; among all, only NGC~4344 is classified by NED as a SB0, but
our visual inspection of the SDSS images of this galaxy reveals no bars).
It means that we have tried to exclude resonance mechanisms of ring
formation in the galaxies under consideration. Some nuclear bars found
by us in the sample galaxies during this works are not able to affect the 
large-scale structure of the galaxies sufficiently to form outer rings, as 
we argued earlier \cite{we2010}. We conclude that the most
probable mechanism to form the outer starforming rings in the S0 galaxies
of the sample is galaxy interaction in its various forms. In the galaxies
NGC~7808, NGC~6340, NGC~6028, UGC~4599, IC~522 we observe perhaps the
destruction of a satellite by tidal forces from the large host galaxy
\cite{helmi}. This conclusion is implied by the inhomogeneous
brightness distributions along the rings in these galaxies and their
appearances as tightly wound spirals. The ring formation in NGC~809, NGC~934,
NGC~4344, NGC~4513, UGC~5936, MCG~11-22-15, NGC~6534, and NGC~446 was
probably the result of a vertical satellite impact onto the central part of the
galaxies -- they are the Carthwell-like cases. The ring symmetry degree
gives evidence for almost central impact in  NGC~809, NGC~934, NGC~4513,
UGC~5936, and NGC~446 and slightly offset one in NGC~4344 and MCG~11-22-15.
Such collisions provoke ringlike density waves in the galactic disks
which are described by Athanassoula \& Bosma \cite{atha} and simulated in
particular by Mapelli et al. \cite{mapelli}. The star formation in the ringlike
density waves was inevitable if a sufficient amount of gas was present
in a disk before the impact.

Some comments about particular galaxies.

NGC~934 and NGC~4513 have the rings which are much brighter in the UV than
in the optical bands. We can treat these rings as close relatives of
the Type I XUV-disks in spiral galaxies \cite{galex};
perhaps it is the earliest stage of the impact-produced rings.

The ring in NGC~4344 is very knotty; it may be classified as the ring of
RE-type according to Athanassoula \& Bosma \cite{atha} criteria. The reason for
such fragmentation may be the absence of stabilizing effect from a central 
mass concentration -- we have already noted that the galaxy lacks bulge.

As for the cases of NGC~4513 and UGC~4599, where there are two starforming
rings seen in the optical bands but only one seen in the UV, we would involve
effects of secular evolution to explain such structures. We would like to note
that the extended UV-disks in the centers of galaxies where in the optical band
we see only narrow starforming rings represent a rather frequent phenomenon.

Perhaps, secular evolution is also responsible for the blue `middle circle'
seen against the reddish broad rings in UGC~5936 and MCG~11-22-15.

And finally, the inclined ring in NGC~252, with its inhomogeneous and asymmetric
optical appearance, must be the result of accreting a satellite from the 
inclined orbit \cite{atha}.

\section{Conclusions}

By using the public data of the GALEX imaging survey of nearby galaxies, we
have compiled a list and have fulfiled surface photometry of 14 lenticular
galaxies with the UV rings (or, in other words, with the outer starforming rings).
The optical colour profiles and the mean colours are calculated for the rings
by using the residual images after subtracting the large-scale photometric
models constructed from the SDSS $g-$ and $r-$-images.

We have discussed the possible ways to form the outer rings possessing the 
morphological and photometric properties similar to those found by us in the 
sample galaxies. Dynamical simulations presented in the literature show a good
agreement with our data if the concept of impact or tidal nature of the rings
is accepted.

The work is based on the public data of the sky surveys GALEX and SDSS.
Funding for the Sloan Digital Sky Survey (SDSS) and SDSS-II has been provided by the Alfred P. Sloan Foundation, the Participating Institutions, the National Science Foundation, the U.S. Department of Energy, the National Aeronautics and Space Administration, the Japanese Monbukagakusho, and the Max Planck Society, and the Higher Education Funding Council for England. The SDSS Web site is http://www.sdss.org/.
The SDSS is managed by the Astrophysical Research Consortium (ARC) for the Participating Institutions. The Participating Institutions are the American Museum of Natural History, Astrophysical Institute Potsdam, University of Basel, University of Cambridge, Case Western Reserve University, The University of Chicago, Drexel University, Fermilab, the Institute for Advanced Study, the Japan Participation Group, The Johns Hopkins University, the Joint Institute for Nuclear Astrophysics, the Kavli Institute for Particle Astrophysics and Cosmology, the Korean Scientist Group, the Chinese Academy of Sciences (LAMOST), Los Alamos National Laboratory, the Max-Planck-Institute for Astronomy (MPIA), the Max-Planck-Institute for Astrophysics (MPA), New Mexico State University, Ohio State University, University of Pittsburgh, University of Portsmouth, Princeton University, the United States Naval Observatory, and the University of Washington.
During the data analysis we have
used the Lyon-Meudon Extragalactic Database (HYPERLEDA) supplied by the
LEDA team at the CRAL-Observatoire de Lyon (France) and the NASA/IPAC
Extragalactic Database (NED) which is operated by the Jet Propulsion
Laboratory, California Institute of Technology, under contract with
the National Aeronautics and Space Administration. 
Our study of the rings in lenticular galaxies is supported by the
RFBR grant number 10-02-00062a.

\clearpage
\centerline{\large \bf Appendix}

\noindent
The colour $g-r$ maps of the residual images of the sample galaxies after
subtracting the model `bulge$+$disks' images from the original SDSS images:\\
1st row -- IC 522, MCG 11-22-15, NGC 252;\\
2nd row -- NGC 446, NGC 809, NGC 934;\\
3rd row -- NGC 4344, NGC 4513, NGC 6028;\\
4th row -- NGC 6340, NGC 6534, NGC 7808;\\
5th row -- UGC 4599, UGC 5936.\\

\begin{figure*}
\includegraphics[width=0.32\hsize]{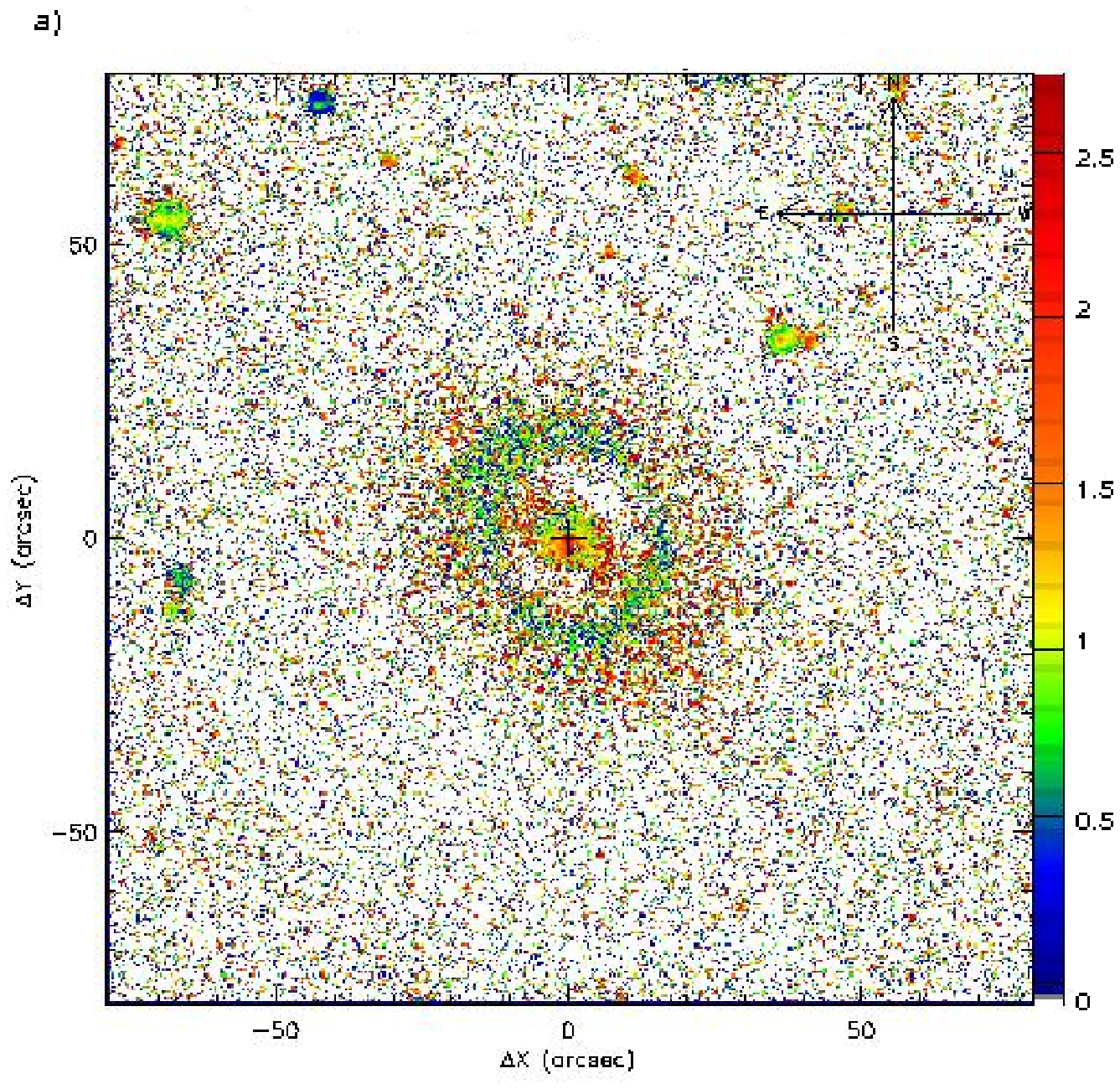}
\includegraphics[width=0.32\hsize]{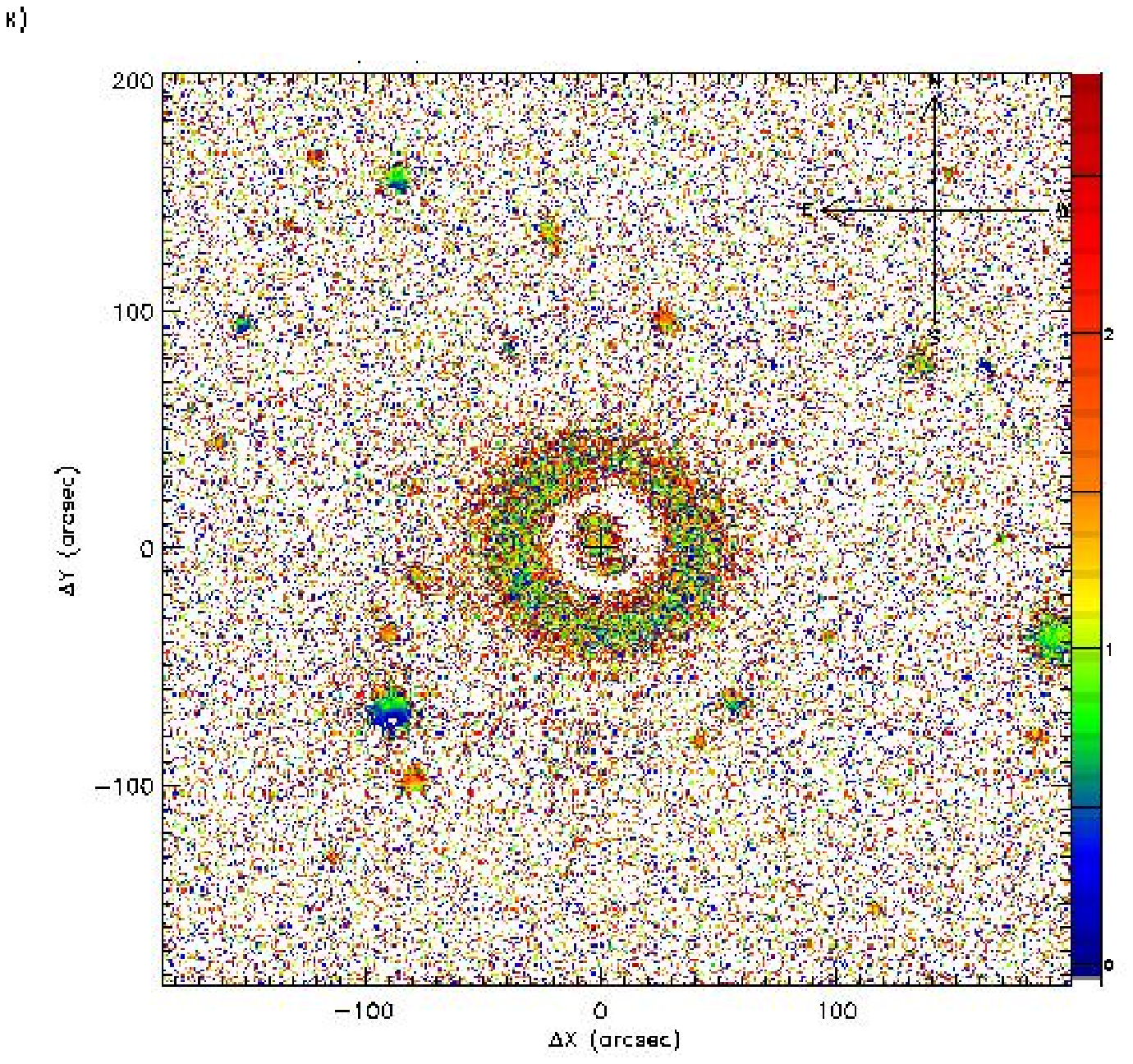}
\includegraphics[width=0.32\hsize]{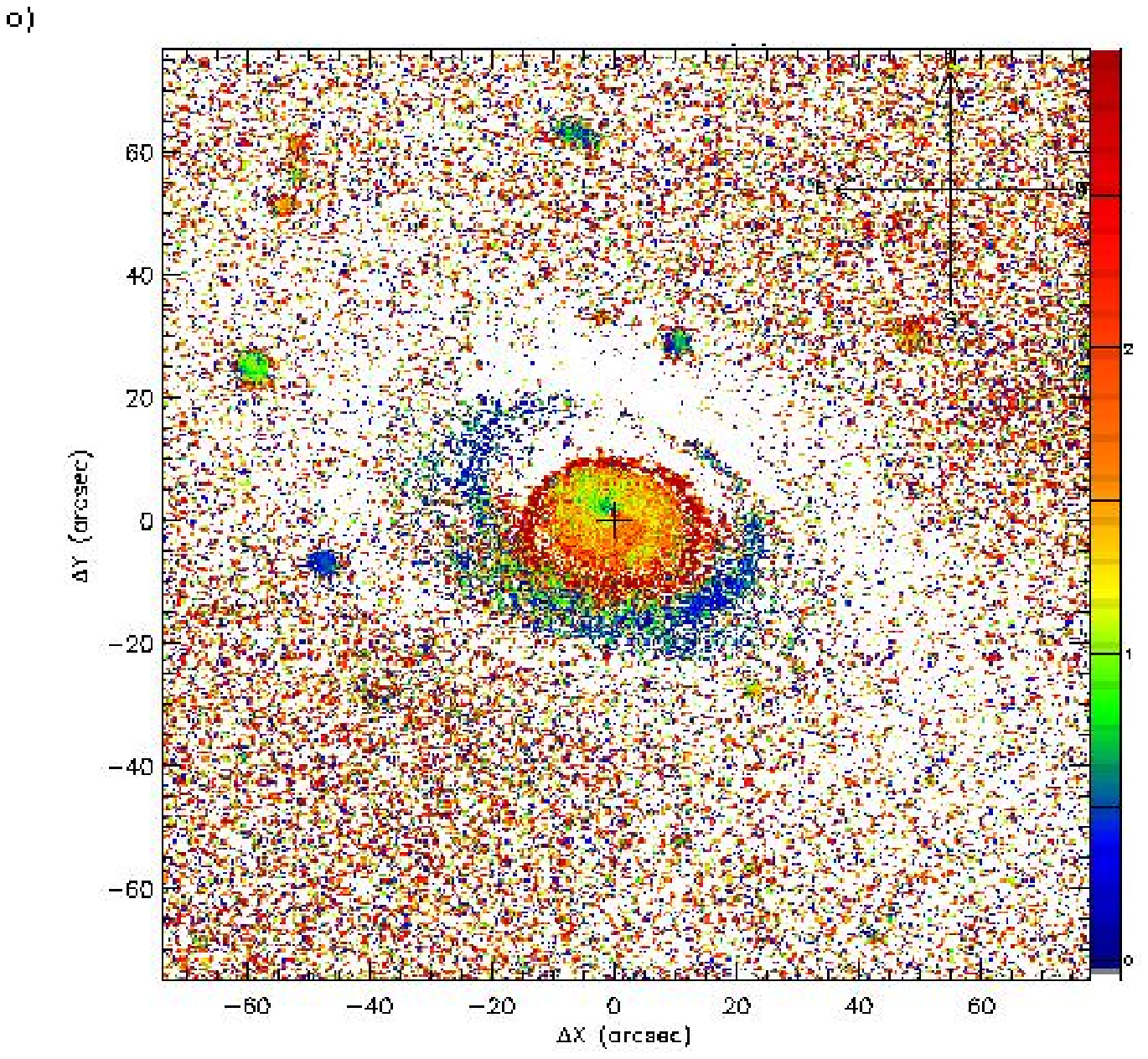}\\
\includegraphics[width=0.32\hsize]{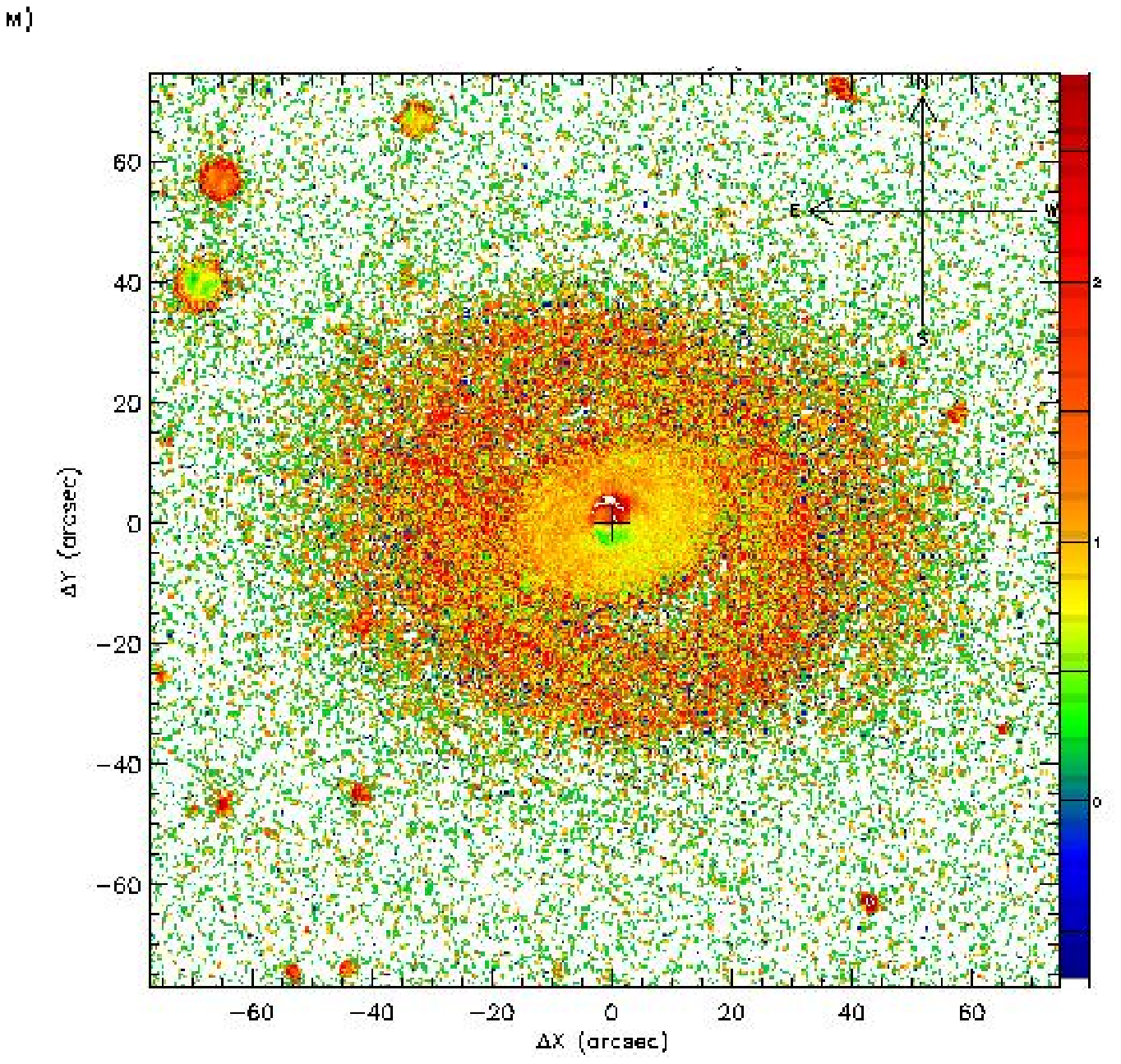}
\includegraphics[width=0.32\hsize]{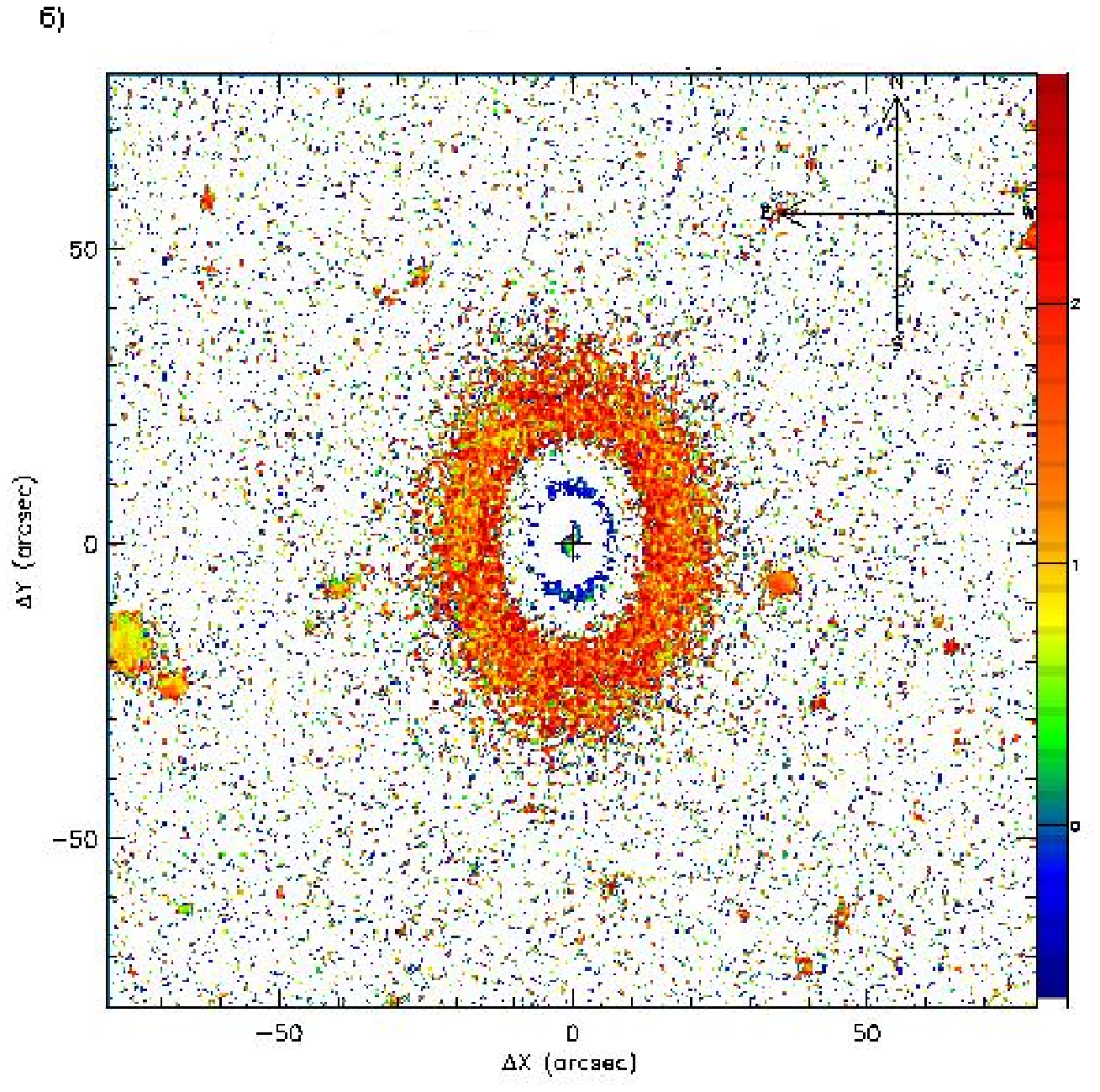}
\includegraphics[width=0.32\hsize]{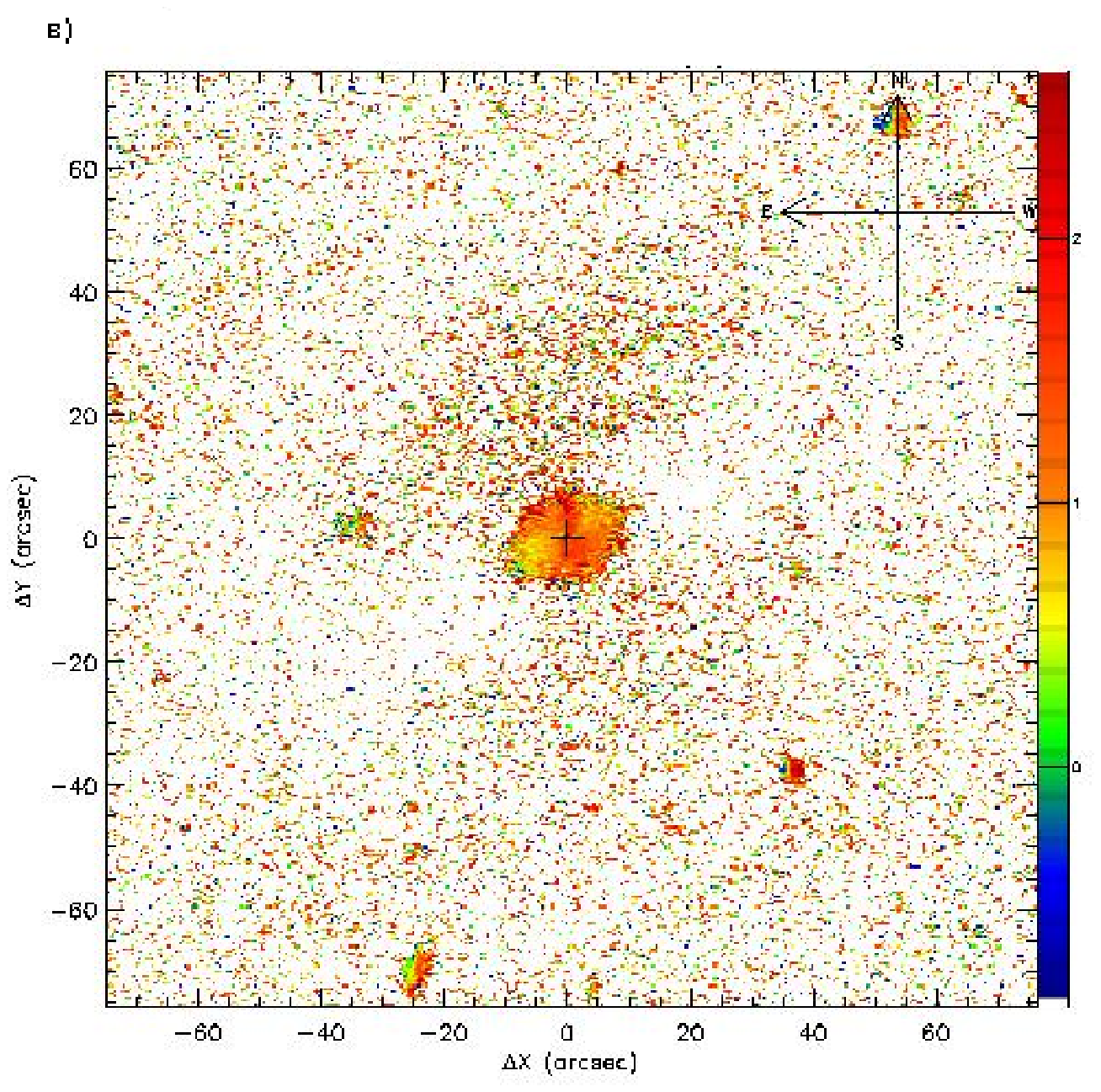}\\
\includegraphics[width=0.32\hsize]{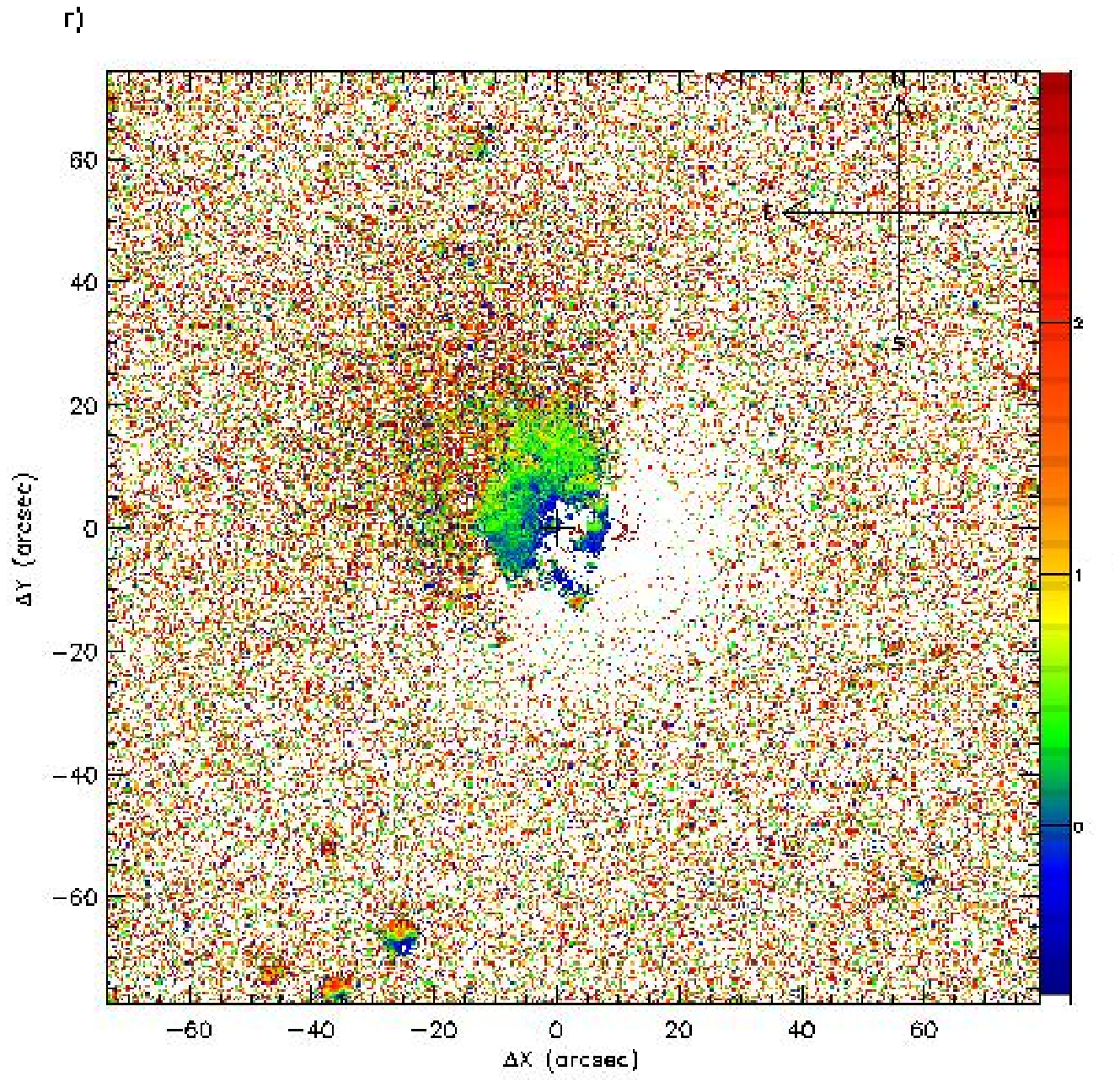}
\includegraphics[width=0.32\hsize]{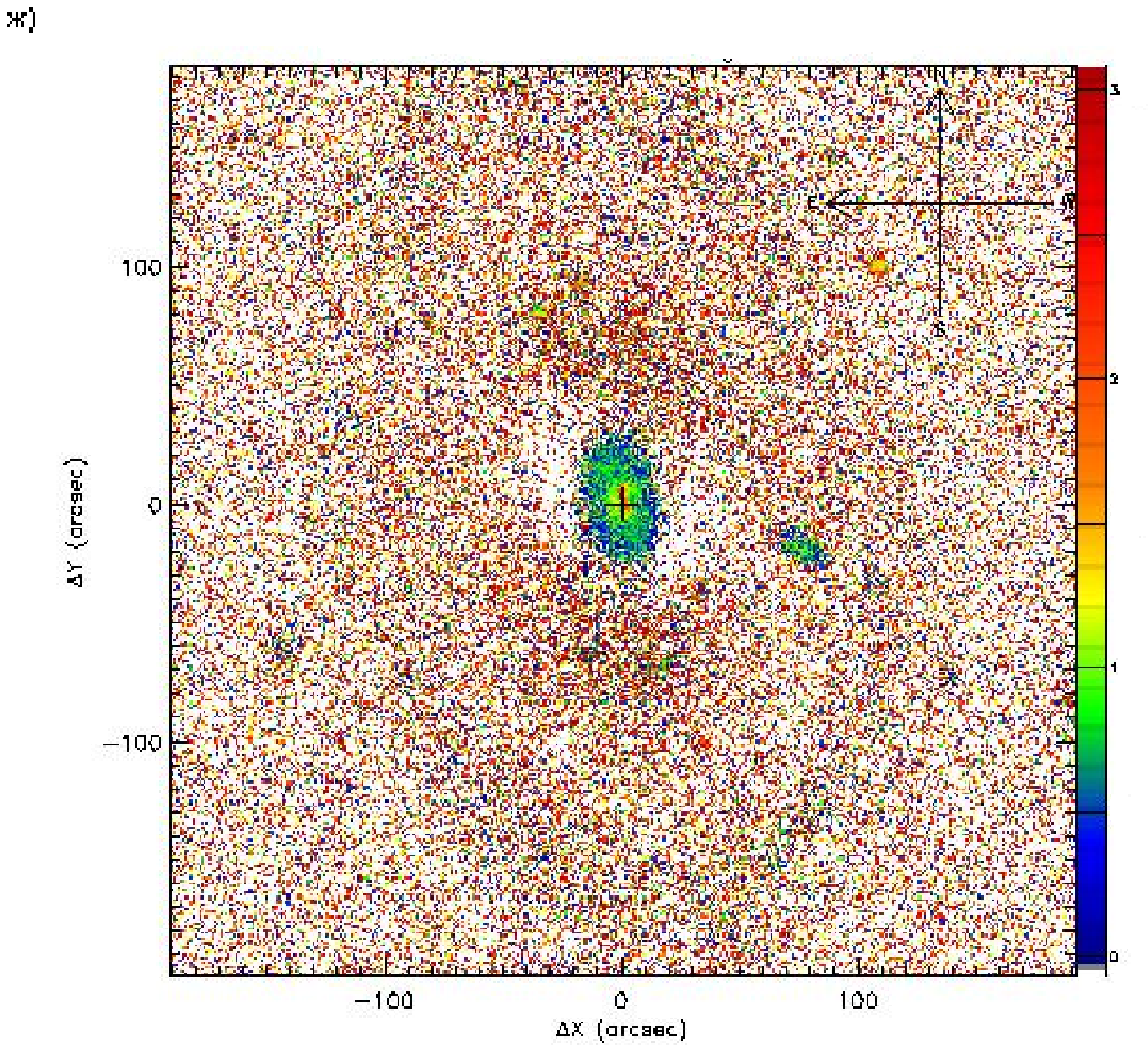}
\includegraphics[width=0.32\hsize]{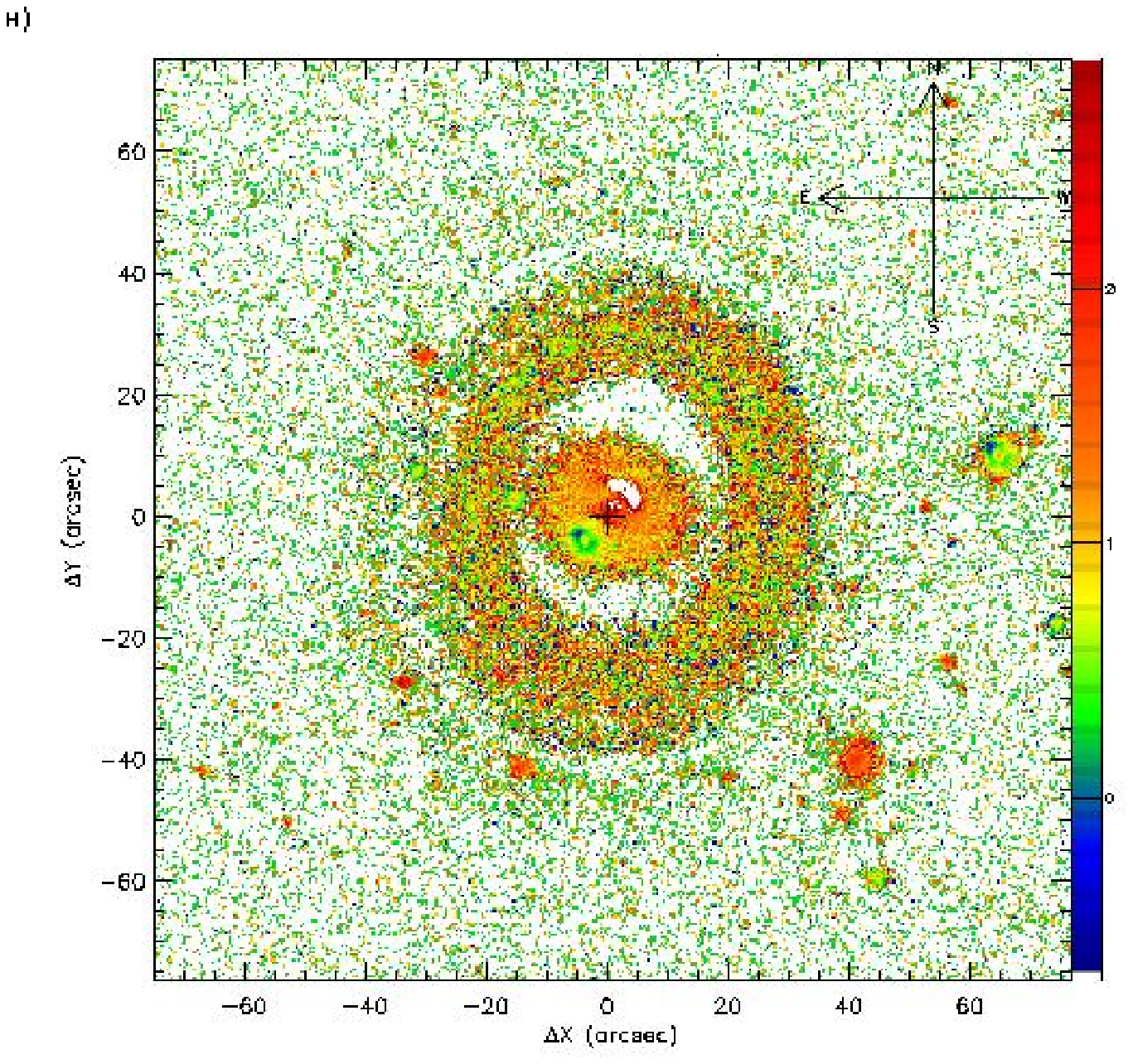}\\
\includegraphics[width=0.32\hsize]{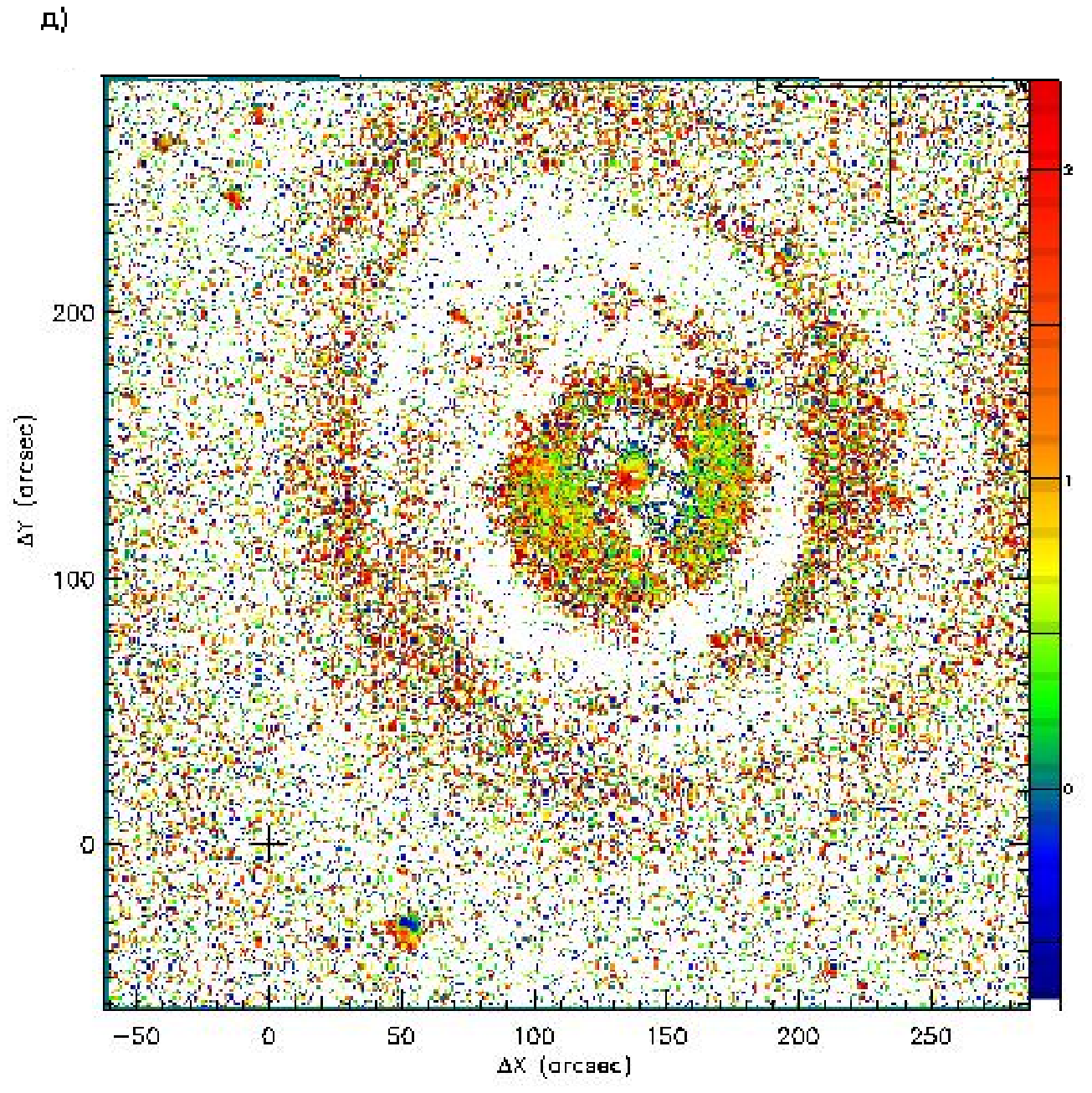}
\includegraphics[width=0.32\hsize]{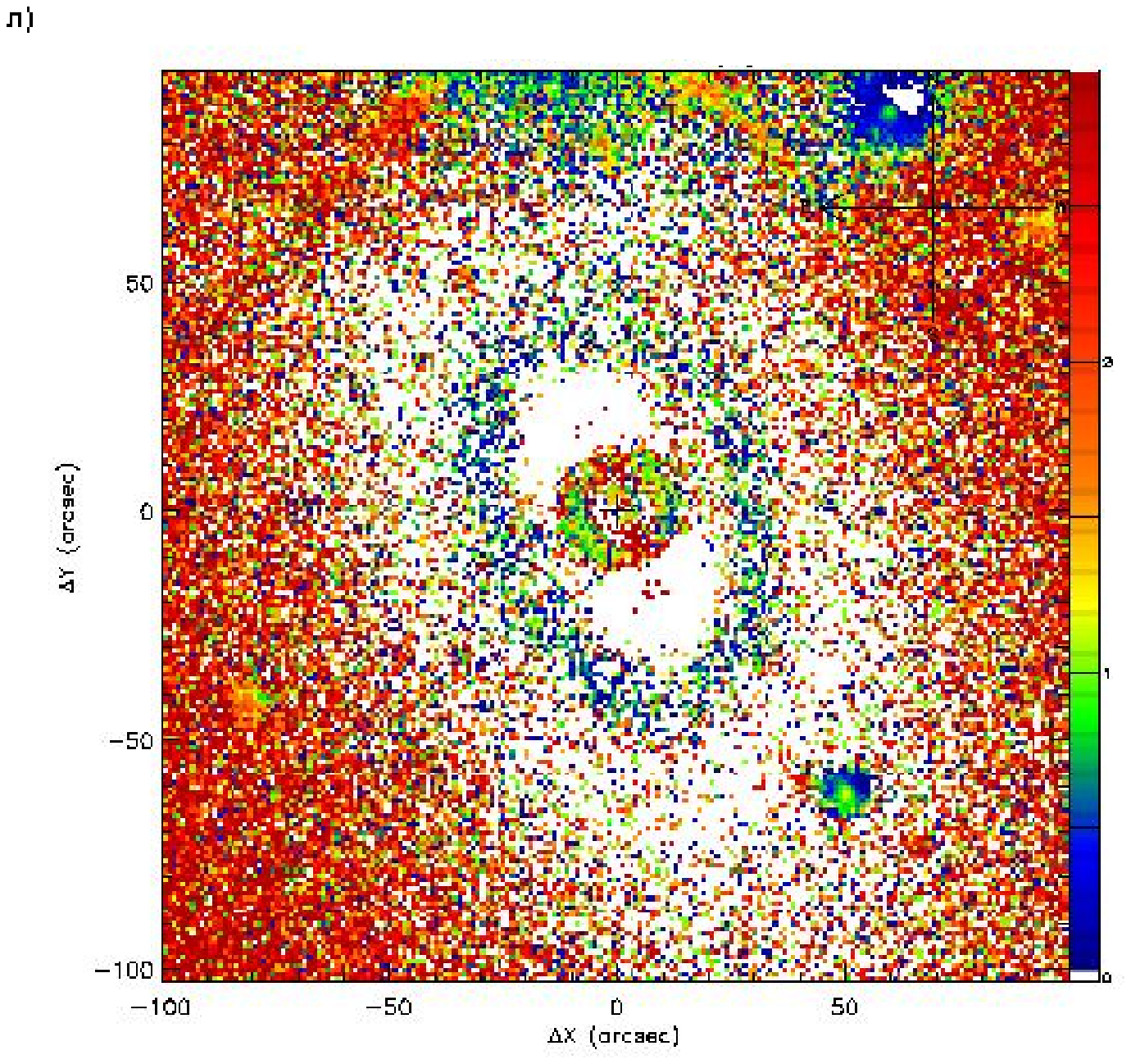}
\includegraphics[width=0.32\hsize]{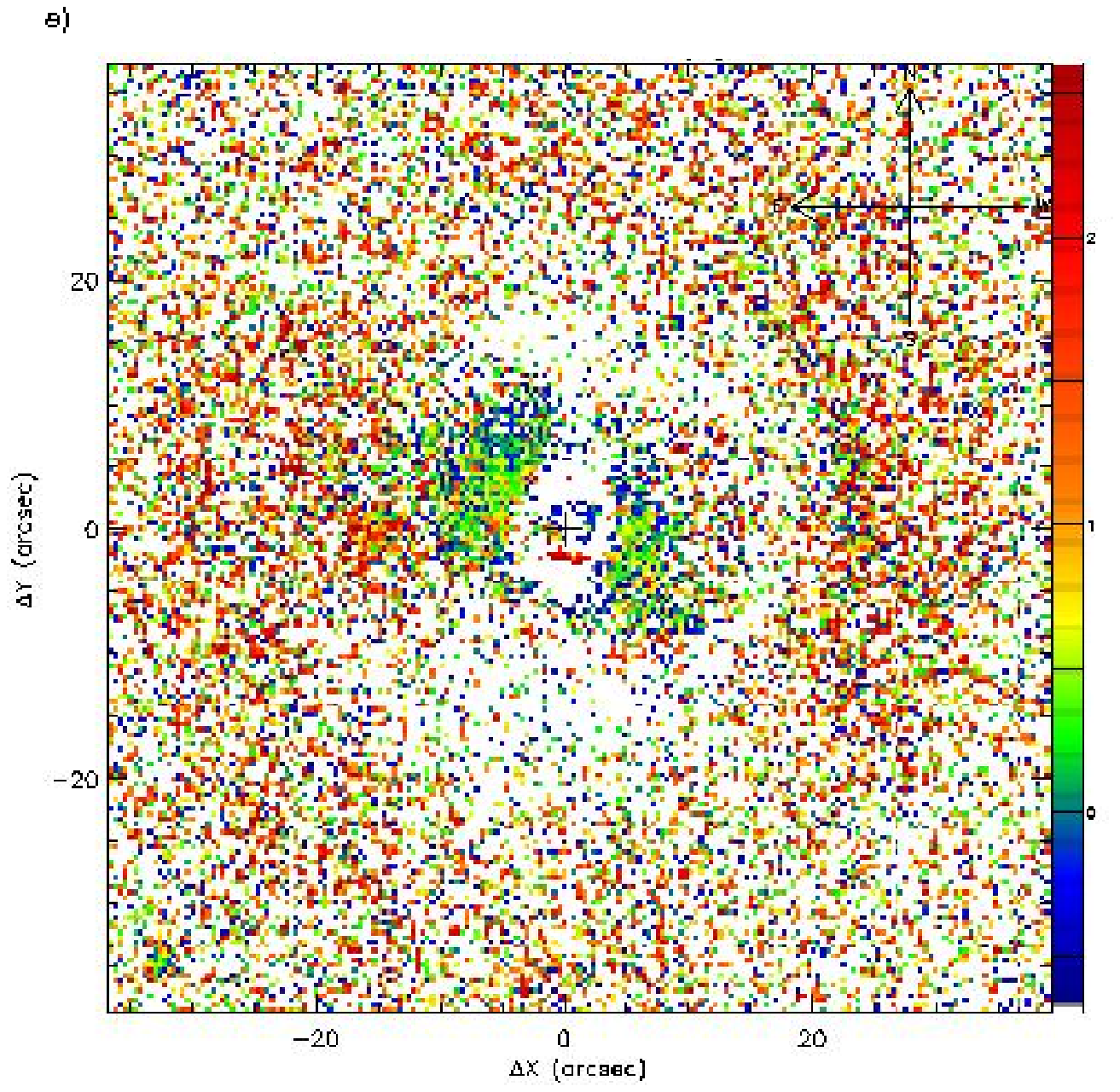}\\
\includegraphics[width=0.32\hsize]{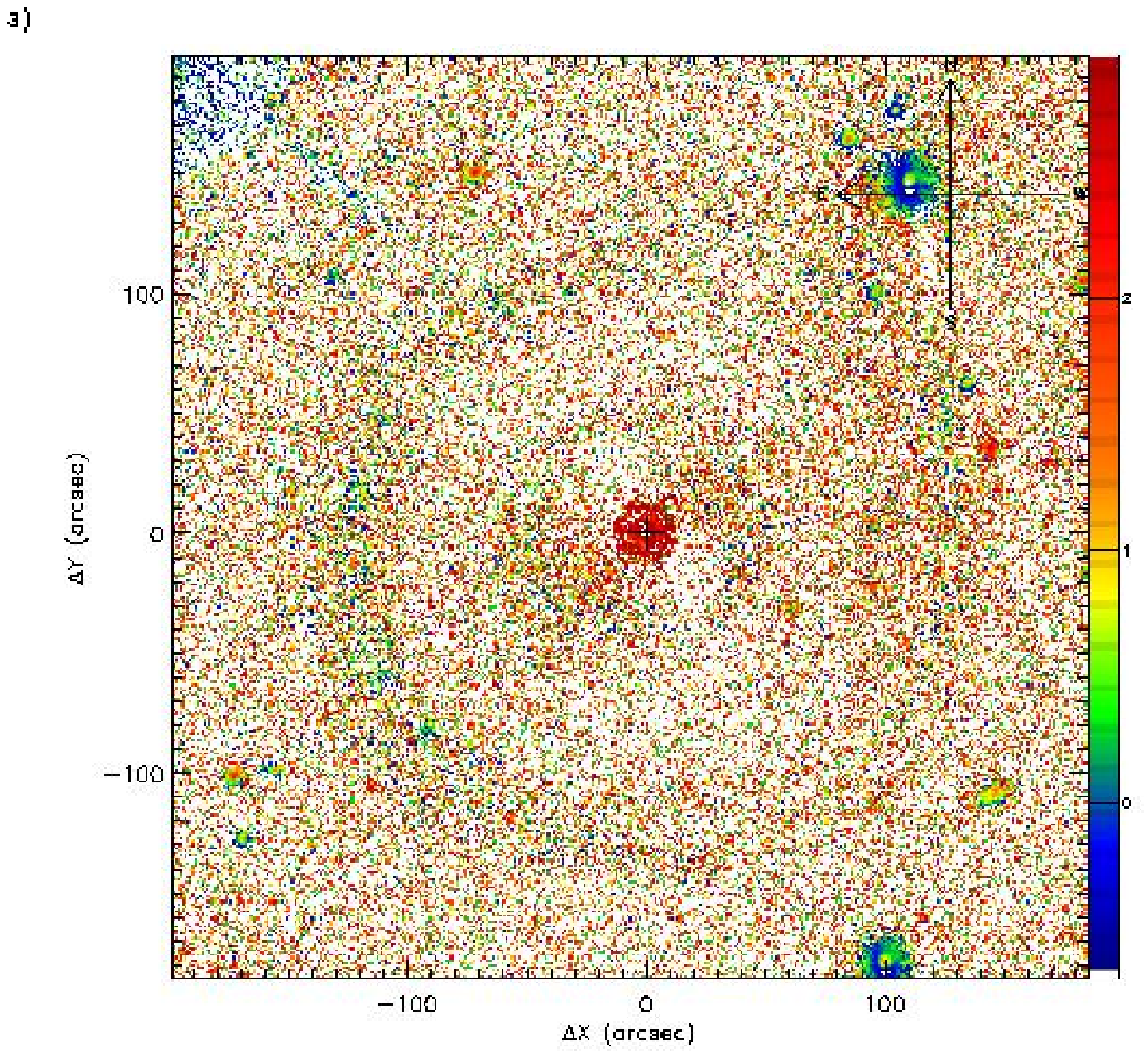}
\includegraphics[width=0.32\hsize]{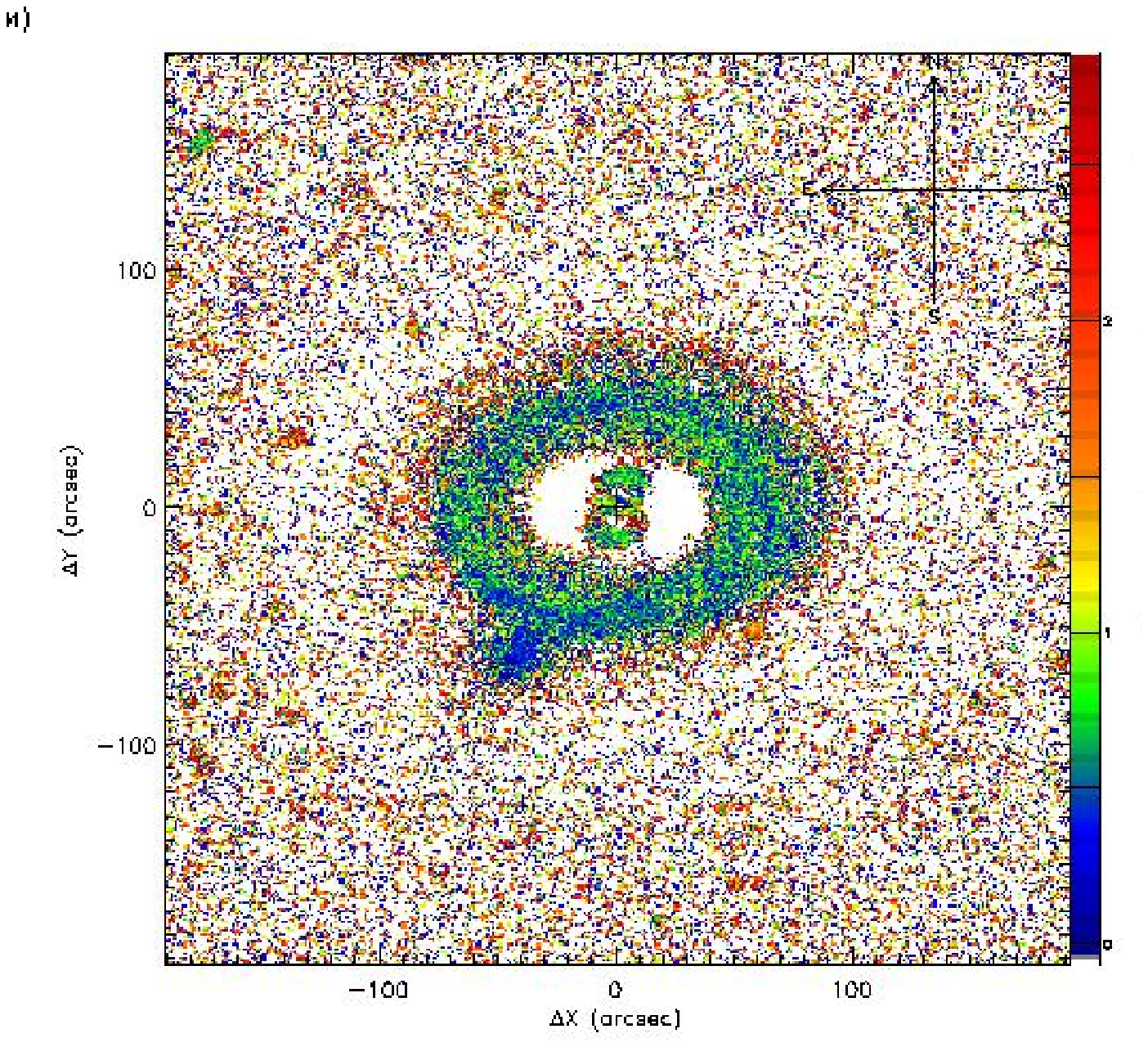}
\caption{
}
\end{figure*}

\end{document}